\documentclass[useAMS, usenatbib]{mn2e}
\usepackage{amsmath,graphicx}

\def\be{\begin{equation}}
\def\ee{\end{equation}}
\def\ba{\begin{eqnarray}}
\def\ea{\end{eqnarray}}
\def\L{\left[\Lambda\right]}
\def\T{\left[\tau\right]}
\def\H{\left[h\right]}
\def\N{\left[n\right]}
\def\A{\left[a\right]}
\def\SQ{\sqrt{\frac{5}{8}e^2 + \chi^2}}
\def\SQEQ{\sqrt{\frac{5}{8}e^2_{\rm eq} + \chi^2}}

\def\go{\mathrel{\raise.3ex\hbox{$>$}\mkern-14mu
             \lower0.6ex\hbox{$\sim$}}}

\def\lo{\mathrel{\raise.3ex\hbox{$<$}\mkern-14mu
             \lower0.6ex\hbox{$\sim$}}}

\begin{document}
\title[Aero-Resonant Migration]
{Aero-Resonant Migration}
\author[Natalia I. Storch and Konstantin Batygin]
{Natalia I. Storch$^1$ and Konstantin Batygin$^2$\\
$^1$ TAPIR, Walter Burke Institute for Theoretical Physics, Mailcode 350-17, Caltech, Pasadena, CA 91125, USA\\
$^2$ Division of Geological and Planetary Sciences, Caltech, Pasadena, CA 91125, USA}

\label{firstpage}
\maketitle

\begin{abstract}
The process of planet conglomeration, which primarily unfolds in a geometrically thin disk of gas and dust, is often accompanied by dynamical excitation of the forming planets and planetesimals. The ensuing orbital crossing can lead to large-scale collisional fragmentation, populating the system with icy and rocky debris. In a gaseous nebula, such leftover solid matter tends to spiral down towards the host star due to aerodynamic drag. Along the way, the inward drifting debris can encounter planets and gravitationally couple to them via mean-motion resonances, sapping them of their orbital energy and causing them to migrate. Here, we develop a simple theory for this migration mechanism, which we call ``Aero-Resonant Migration'' (ARM), in which small planetesimals ($10 \mathrm{m} \lo s \lo 10$km) undergo orbital decay due to aerodynamic drag and resonantly shepherd planets ahead of them. Using a combination of analytical calculations and numerical experiments, we show that ARM is a robust migration mechanism, able to significantly transport planets on timescales $\lo 1$ Myr, and present simple formulae for the ARM rate.
\end{abstract}


%%%%%%%%%%%%%%%%%%%%%%%%%%%%%

\section{Introduction}

Classical planetary formation theory, developed before the wide-spread detection of exoplanets, has historically treated planetary system formation as an inherently static process. By invoking the Minimum Mass Solar Nebula (MMSN) -- a \textit{lower limit} to the amount of material needed within the protoplanetary disk in order to form all the planets (Hayashi 1981, Weidenschilling 1977) -- Solar System formation scenarios generally assumed that the amount of solid matter in the system has remained largely unchanged and that its current radial distribution closely resembles the primordial one. Under this assumption, planets never strayed far from their birthplaces. 

One of the key realizations that arose from exoplanet characterization is that planetary systems are \textit{not} static. A large sample of planets on eccentric orbits (e.g. Wright et al. 2011) hints at dynamical restructuring of orbital architectures (Chatterjee et al. 2008; Juri\'c \& Tremaine 2008; Nagasawa \& Ida 2011; Beaug\'e \& Nesvorn\'y 2012), while multi-planet systems in which planets sit close to, though not necessarily in, mean-motion resonances (MMRs; e.g. Fabrycky et al. 2014; Veras \& Ford 2012) suggest a history of convergent migration in which either planets fail to capture in MMR (e.g. Matsumoto et al. 2012; Baruteau \& Papaloizou 2013; Goldreich \& Schlichting 2014), or do capture into MMR and later break out due to various effects (e.g. Rein 2012; Batygin \& Morbidelli 2013; Chatterjee \& Ford 2015). Indeed, the predominance of compact systems in which several planets are packed into tight orbits with semi-major axes smaller than that of Mercury (e.g. Batalha et al. 2013) increasingly points to the notion that the Solar System is but one outcome of a physically diverse process of planet formation. Moreover, modern models of Solar System formation, too, invoke considerable redistribution of matter during the Solar System's infancy, suggesting that its orbital architecture has been dynamically sculpted (see e.g. Morbidelli et al. 2012 for a review). Finally, the existence of ultra-short-period planets (Sanchis-Ojeda et al. 2014), which reside inside the star's putative magnetospheric cavity, likely also necessitates migration.

To date, a variety of planetary migration mechanisms have been considered in the literature. These include smooth migration through the gaseous protoplanetary disk -- type I migration for lower-mass planets (Goldreich \& Tremaine 1979, Ward 1997), type II for planets with mass comparable to that of Jupiter (Lin \& Papaloizou 1986) and type III for intermediate, Saturn-mass, planets (Masset \& Papaloizou 2003); migration through dynamical interaction with a disk of remnant planetesimals after gas dispersal, as in the Nice Model (Tsiganis et al. 2005, Batygin \& Brown 2010, Levison et al. 2011); and late-type migration through dynamical interactions between planets, or with a distant perturber, collectively known as high-eccentricity migration (e.g. Beaug\'e \& Nesvorn\'y 2012, Naoz 2016). The relative dominance and specific roles of each of these processes remains an area of active research.

In theory, the dominant planet sample, which consists of super-Earths with masses less than or comparable to Neptune's, should be primarily subject to type I migration, in which a slight angular momentum imbalance between the leading and trailing arms of a spiral wave raised by the planet causes it to exchange angular momentum with the disk. However, detailed numerical simulations have shown that both the magnitude and the direction of type I migration are extremely sensitive to the entropic structure of the protoplanetary disk, making it difficult to deterministically understand its consequences (e.g. Bitsch \& Kley 2011; Bitsch et al. 2014, 2015). Here we show that this population of planets ($2M_\oplus \lo M_p \lo 30 M_\oplus$) is also subject to a different smooth migration mechanism, which we call ``Aero-Resonant Migration'', or ARM. In this mechanism, small planetesimals undergoing inward migration due to aerodynamic drag (AD) from the gaseous disk become captured into exterior first-order mean-motion resonances with the planet. Trapped in resonance, they continue experiencing AD and act as an energy sink for the planet. If the total mass of trapped planetesimals is appreciable, the planet migrates inward. 

A proof of concept for ARM was first presented by Batygin \& Laughlin (2015). Here we use a combination of analytical and numerical methods to further develop the theory of ARM, and outline the regime in which it dominates. In section II, we present the setup and our analytical treatment of the problem. In section III we present the results of numerical experiments that serve to validate the conclusions of section II. We conclude briefly in section IV. 

\section{ARM: Analytical Treatment}

\subsection{Model Setup and Definitions}

We consider a gaseous protoplanetary disk around a central star of mass $M_\star$, with a surface density profile $\Sigma$ (Mestel 1963) such that
\be
\Sigma = \Sigma_0\left(\frac{a_0}{a}\right),
\ee
\noindent where $a$ denotes semi-major axis, and $\Sigma_0$ is the disk surface density at reference semi-major axis $a_0$. Due to radial pressure support, the gas disk is marginally sub-Keplerian, and we define a parameter $\chi$ as
\be
{\bf v}_{\rm gas} =v_{\rm K} \sqrt{1-3\frac{c_{\rm s}^2}{v_{\rm K}^2}}{\bf \hat{\phi}} \equiv v_{\rm K} (1-\chi){\bf \hat{\phi}},
\ee
\noindent where $c_s$ and $v_{\rm K}$ are the sound and Keplerian speeds, respectively, and ${\bf \hat{\phi}}$ is the azimuthal unit vector. Typically, $\chi \sim 0.001 - 0.01$. 

Embedded in the gaseous disk is a planet of mass $m_p$, at semi-major axis $a_p$, and a population of small planetesimals of mass $m$, radius $s$, whose radial density profile initially follows that of the gaseous disk. The total mass of the planetesimal disk is a factor $f$ smaller than that of the gas disk; typically, $f \sim 0.01$. For simplicity, we treat $m$ and $s$ as adjustable parameters, keeping in mind that the real disk has a distribution of planetesimals of varying sizes.

We restrict the planetesimals to have $s \go 10$ m, such that their hydrodynamic Reynolds number substantially exceeds unity (Malhotra 1993). Note that this assumption is only limiting if the mass-fraction of solids in the $10~\mathrm{cm}\lo s \lo 10$ m range is large. Under this assumption, the planetesimals experience aerodynamic drag (AD) with an acceleration $\mathbf{a}_{\mathrm{drag}}$ that is quadratic in velocity and independent of the gas viscosity (Landau \& Lifshitz 1959),
\be
\mathbf{a}_{\mathrm{drag}}=-\frac{\pi C_d}{2m}s^2\rho_{\mathrm{gas}}v_{\mathrm{rel}}\mathbf{v}_{\mathrm{rel}}.
\label{adrag}
\ee
\noindent Here, $\mathbf{v}_\mathrm{rel} \equiv \mathbf{v}-\mathbf{v}_\mathrm{gas}$ is the relative velocity between the gas and the planetesimals, $\rho_\mathrm{gas}$ is the gas volume density and $C_d \simeq 0.5$ is the drag coefficient. 

The effect of the AD acceleration is to cause the planetesimals to drift inward, and to damp their eccentricities and inclinations. Since in this work we are concerned solely with first order resonances, which cannot excite inclinations, we set all planetesimal inclinations to zero. Then, to first order in eccentricity, the migration and eccentricity damping rates can be expressed as 
\begin{align}
\frac{1}{a}\frac{da}{dt} &= - \frac{2}{\tau}\chi\SQ, \label{eq:adot}\\
\frac{1}{e}\frac{de}{dt} &= - \frac{1}{\tau}\SQ, \label{eq:edot}
\end{align}
\noindent where
\be
\tau = \left(\frac{\pi C_{\rm D}}{2 m} s^2 \rho_{\rm gas} v_{\rm K}\right)^{-1} \label{tau}
\ee
\noindent is the characteristic damping timescale. Note that, with the above assumptions for the disk density, we have $\tau \propto a^{5/2}$.

As those planetesimals with initial semi-major axes greater than $a_p$ drift inward, they encounter and become captured into an exterior $k:k-1$ mean-motion resonance with the planet. For the remainder of this section, we consider the consequences of this capture. Readers that are not interested in the mathematical details are encouraged to skip the next two subsections and proceed directly to subsection 2.4.

\subsection{Effect of a single planetesimal in resonance}

Consider the resonant interaction between the planet, denoted with subscript $1$, and a single planetesimal, denoted with subscript $2$, with mass $m_2 \ll m_1$ trapped in resonance. Since $m_2 \ll m_1$ we assume that $e_1 = 0$. 

To evaluate the planet migration rate $\dot{a}_1$ induced by this single particle, we use the Hamiltonian formalism of Batygin \& Morbidelli (2013). We start by defining Poincare action-angle variables
\be
\Lambda = m\sqrt{GMa}, \quad \lambda = \mathcal{N}+\varpi
\ee
\be
\Gamma = \Lambda\left(1-\sqrt{1-e^2}\right)\approx \Lambda e^2/2, \quad \gamma = -\varpi,
\ee
\noindent where $\mathcal{N}$ is the mean anomaly, $\varpi$ is the longitude of periastron.
We work to first order in eccentricity. Then the Hamiltonian consists of a Keplerian and a resonant piece, given by
\begin{align}
H_{\rm K} &= -\frac{G^2 M^2 m_1^3}{2\Lambda_1^2} - \frac{G^2 M^2 m_2^3}{2\Lambda_2^2},
\label{HKep} \\
H_{\rm res} &= - \frac{G^2 M m_1 m_2^3}{\Lambda_2^2} f_{\rm res}^{(2)}\sqrt{\frac{2\Gamma_2}{\Lambda_2}}\cos\left(k\lambda_2 - (k-1)\lambda_1 + \gamma_2\right).\label{Hres} 
\end{align}
\noindent Here $f_{\rm res}^{(2)}$ is a positive quantity of order unity that weakly depends on the ratio $a_1/a_2$.

Because we are working to first order in eccentricity, we may expand these Hamiltonians around the nominal resonance location. All quantities computed at the nominal resonance are denoted with $[~]$. After much manipulation (see Batygin \& Morbidelli 2013), we arrive at the Hamiltonian
\be
H = \eta \Phi_2 + \beta\sqrt{2\Phi_2}\cos\phi_2,
\label{hammy}
\ee
\noindent where
\begin{align}
\eta &\equiv 3\Bigl(\left[h\right]_1 (k-1)\Psi_1 - \left[h\right]_2 k \Psi_2\Bigr), \label{eta} \nonumber \\
\Psi &\simeq \Lambda = m\sqrt{G M a}, \nonumber \\
\Phi &= \Gamma \simeq \frac{1}{2}\Lambda e^2 \simeq \frac{1}{2} \L e^2, \\
\left[h\right] &\equiv \left[n\right]/\left[\Lambda\right] = 1/(m\left[a\right]^2), \nonumber \\
\phi_2 &\equiv k \lambda_2 - (k-1)\lambda_1 + \gamma_2, \nonumber \\
\beta &\equiv -\frac{G^2 M m_1 m_2^3}{\left[\Lambda\right]_2^2} \frac{f_{\rm res}^{(2)}}{\sqrt{\left[\Lambda\right]_2}}.\nonumber
\end{align}

\noindent Here $\eta$ is a measure of proximity to nominal resonance, $\phi_2$ is the resonant angle, and $\beta$ is the strength of the resonance. Note that $\beta < 0$. Importantly, the Hamiltonian of Eq. (\ref{hammy}) is only an appropriate model for resonant dynamics at very low eccentricity, and does not capture any exotic resonant behavior. This is, however, sufficient for the problem at hand.

From this point on, we will drop the subscript $2$ for simplicity but retain the subscript $1$ wherever appropriate. Further defining for convenience the Cartesian variables
\be
x = \sqrt{2\Phi}\sin\phi \quad y = \sqrt{2\Phi}\cos\phi,
\ee
\noindent we then arrive, via Hamilton's equations, at very simple equations of motion:
\begin{align}
\dot{x} &= \beta + \eta y, \label{eq:xdot}\\
\dot{y} &= -\eta x. \label{eq:ydot}
\end{align}

It is now necessary to account for the effect aerodynamic drag. The effect of AD is to diminish the phase space area occupied by the particle's trajectory to the point where the time derivatives of all the variables are null. Since semi-major axis decay occurs at a rate slower than eccentricity decay by a factor of $\chi$, we may split this process into two parts. First we derive the equilibrium eccentricity attained by the particle at a fixed semi-major axis (or, equivalently, a fixed proximity parameter $\eta$). Second, we will consider the evolution of $\eta$ due to semi-major axis decay, and find its equilibrium value. 
To first address the question of equilibrium eccentricity, we add to equations (\ref{eq:xdot}) \& (\ref{eq:ydot}) the AD eccentricity damping term of Eq.(\ref{eq:edot}), arriving at
\begin{align}
\dot{x} &= \beta + \eta y - \frac{x}{\tau}\SQ, \\
\dot{y} &= -\eta x - \frac{y}{\tau}\SQ.
\end{align}

We wish to find $e$ and $\phi$ such that $\dot{x}=\dot{y}=\dot{e}=\dot{\phi}=0$. After some manipulation, we find
\begin{align}
e_{\rm eq} &= \frac{1}{\sqrt{\L}}\frac{\beta}{\eta}, \\
\phi_{\rm eq} &= \pi - \frac{1}{\tau\eta}\SQEQ. 
\end{align}
\noindent Note that this solution only yields positive values of eccentricity when $\eta < 0$ (because $\beta < 0$). 

ARMed (get it?) with $e_{\rm eq}$ and $\phi_{\rm eq}$, we can now derive an equation for the evolution of $\eta$ due to $\dot{a}$ from aerodynamic drag. From Eq. (\ref{eq:adot}), we have
\be
\frac{\dot{a}_{\rm AD}}{a} = -\frac{2\chi}{\T}\SQEQ. \label{adotAD}
\ee

\noindent In order to get $\dot{\eta}$, we need $\dot{\Lambda}_1$ and $\dot{\Lambda}_2$. $\dot{\Lambda}_2$ will have two contributions: one that comes directly from the semi-major axis decay due to AD, and the other that comes through the resonant interaction.  $\dot{\Lambda}_1$ will only contribute through the resonant interaction. From Eq. (\ref{adotAD}), we have
\be
\dot{\Lambda}_2\Big|_{\rm AD} = - \L_2 \frac{\chi}{\T}\SQEQ.
\ee
\noindent Then, from Eq. (\ref{Hres}),
\begin{align}
\dot{\Lambda}_1\Big|_{\rm res} &= (1-k) \frac{\beta^2}{\T\eta^2}\SQEQ, \label{L1} \\
\dot{\Lambda}_2\Big|_{\rm res} &= k \frac{\beta^2}{\T\eta^2}\SQEQ.
\end{align}
\noindent Putting this together with Eq. (\ref{eta}),
\begin{multline}
\dot{\eta} = 3\left\{k\H_2\L_2\frac{\chi}{\T} - \bigl(\H_1(k-1)^2+\H_2 k^2\bigr)\frac{\beta^2}{\T \eta^2}\right\}\times \\
\times\SQEQ.
\end{multline}

\noindent Setting the expression inside curly brackets equal to $0$, we find
\be
\eta^2_{\rm eq} = \frac{\beta^2}{\N_2\chi k}\bigl\{\H_1(k-1)^2 + \H_2 k^2\bigr\}.
\ee
Recall that only the $\eta < 0$ solution is valid.

Now that we have $\eta_{\rm eq}$, we can finally calculate the semi-major axis decay rate for the inner planet.\footnote{Recall that $\dot{\Lambda}_1 \propto \beta^2/\eta^2$, and $e_{\rm eq} \propto \beta^2/\eta^2$.} Using the solution from the previous subsection, we find
\be
\begin{split}
\frac{\beta^2}{\eta^2} &= \frac{\N_2 k \chi}{\H_1(k-1)^2 + \H_2 k^2} \\
&= \frac{m\N_2 k \chi}{\A_2^{-2}k^2 + \frac{m}{m_1}\A_1^{-2}(k-1)^2} \\
&\simeq \frac{\L_2 \chi}{k},
\end{split}
\ee
\noindent where in the last step we have assumed $m \ll m_1$. Note that this implies that $e_{\rm eq}$ takes the very simple form
\be
e_{\rm eq}^2 = \frac{\chi}{k}.
\ee
We may now calculate the rate of semi-major axis decay:
\be
\frac{\dot{a}_1}{a_1} = 2\frac{\dot{\Lambda}_1}{\Lambda_1} = -2 \left(\frac{k-1}{k}\right)^{2/3}\frac{m}{m_1}\frac{\chi}{\T}\sqrt{\frac{5}{8}\frac{\chi}{k}+\chi^2}.
\ee

\begin{figure}
\centering
\scalebox{0.43}{\includegraphics{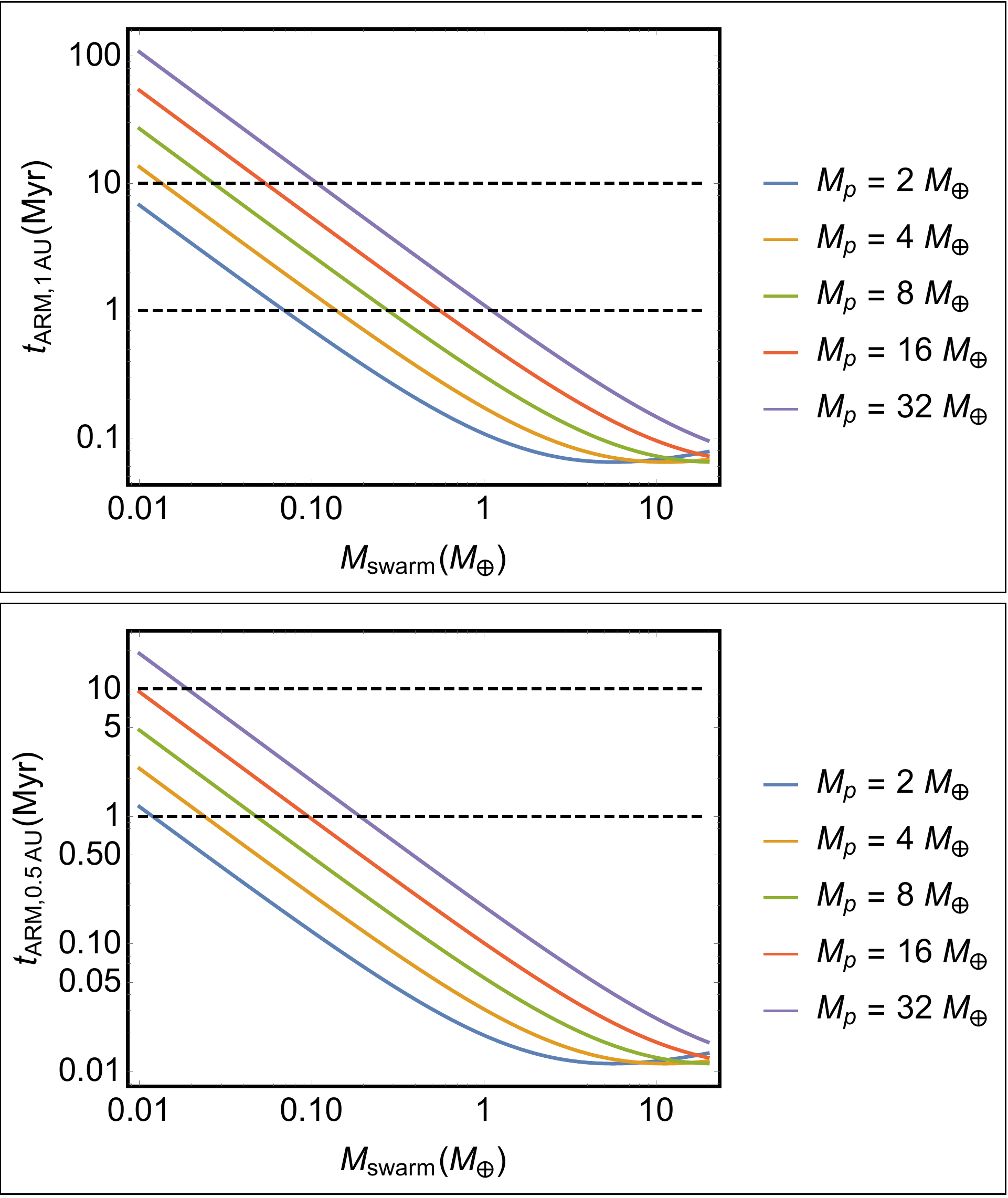}}
\caption{\textit{Top:} Total time $t_{\mathrm{ARM, 1 AU}} \equiv \frac{2}{5}t_\mathrm{ARM}$ that it takes for a planet of mass $M_p$ starting at a semi-major axis of $1$ AU to be pushed into the sun by a swarm of $1$-km planetesimals of mass $M_\mathrm{swarm}$ trapped in the $3:2$ resonance, assuming a MMSN-like protoplanetary disk (see Eq. \ref{tarm} for all nominal parameter values). Evidently, even for the most massive planets, there exists a large range of planetesimal masses for which the migration takes significantly less than $1$ Myr. \textit{Bottom:} Same as top, but starting at $0.5$ AU. Note that just $0.3 M_\oplus$ worth of $1-$km planetesimals is enough to push a $32 M_\oplus$ planet into the sun within 1 Myr.}
\label{analyticals}
\end{figure}

\noindent Note here that $\T$ is evaluated at the nominal resonance location for the outer particle, i.e. at $\A_2 = \left(\frac{k}{k-1}\right)^{2/3}a_1$, and thus overall $\dot{a}_1 \propto a_1^{-3/2}$.

\subsection{Effect of a collection of $N$ planetesimals in resonance}

Having gone through the exercise of deriving the planet's decay rate due to the effect of a single particle trapped in resonance, we can now easily generalize our calculation to include $N$ approximately coorbital particles, each of mass $m$. One caveat is that, in the previous section, since $m \ll m_1$, we freely assumed that the planet eccentricity is null. Here, given that the total mass of the ``swarm'' of particles can in principle be of order the mass of the planet itself, this assumption is not \textit{a priori} valid. Nevertheless, we will make it and check it \textit{a posteriori}. By doing so, in essence, we are assuming that the swarm of particles is azimuthally distributed and they are thus unable to coherently raise the planet's eccentricity. The validity of this assumption will be addressed by the numerical experiments of section III. 

In the simplest approximation, the particles interact with the inner planet but not with each other. At the quantitative level, this approximation cannot be valid, since the particle ``swarm'' has appreciable mass and its self-interaction is not guaranteed to be negligible. As we show below, however, the effects of self-interaction are secondary -- although they affect how long the particles remain in resonance, and modulate the resonant libration amplitudes and eccentricities, the fundamental machinery of ARM is captured by the sole consideration of the resonant planet-particle interactions and the aerodynamic drag the particles experience. Thus, while self-interaction is certainly an important part of any realistic ARM scenario, it is not essential to quantifying the ARM rate. We address the effect of self-interaction further in section $3.2$.

Under the above non-interacting assumption, the analysis for each individual particle proceeds analogously to the analysis presented in the previous subsection, up until the equilibrium proximity parameter needs to be calculated. Because the proximity parameter is affected not only by the semi-major axis evolution of the outer particle, but that of the inner planet as well, the number/total mass of outer particles affects their final equilibrium proximity parameter (since each of them acts on the inner planet). We can quantify this as follows.

\begin{figure*}
\centering
\scalebox{0.4}{\includegraphics{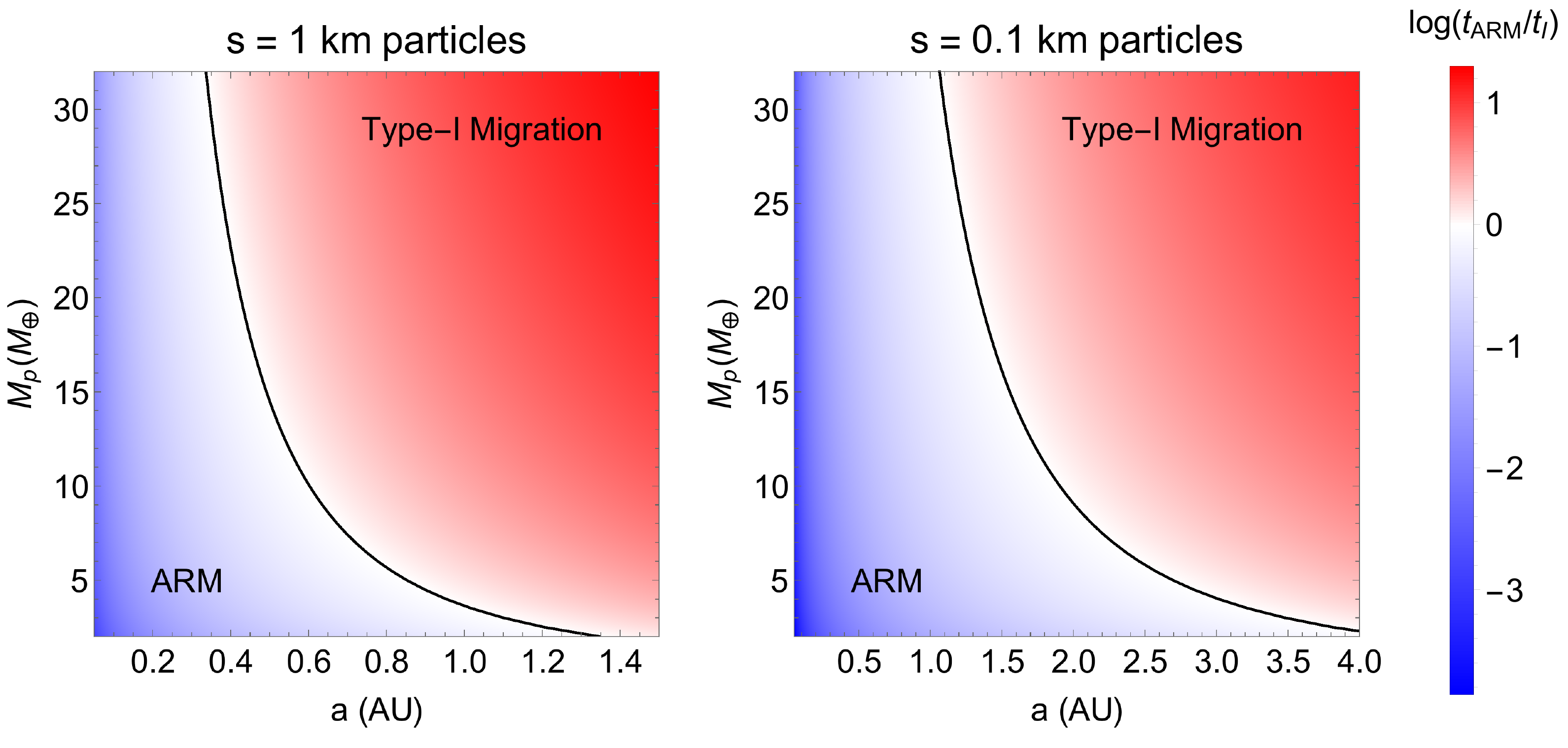}}
\caption{Ratio of the minimum ARM timescale (Eq. \ref{tarm}) to the typical type I migration timescale as a function of planet mass and planet semi-major axis, for $s=1$ km particles (left) and $s=0.1$ km particles (right). The black line traces the curve where the two timescales are equal. Due to the steep dependence of $t_\mathrm{ARM}$ on semi-major axis, it easily dominates in the inner Solar System (inside $\sim0.5$ AU).}
\label{typeIcomp}
\end{figure*}

Due to the fact that $\Lambda_1$ is resonantly affected by each of the outer particles, we must replace the expression for its evolution given in Eq. (\ref{L1}) with
\be
\begin{split}
\dot{\Lambda}_1\Big|_{\rm res} &= \sum_{i=1}^N{-\frac{\partial H_i}{\partial \lambda_1}} \\
&= \sum_{i=1}^N{(1-k)\frac{\beta^2}{\T\eta_i^2}\sqrt{\frac{5}{8}e_i^2 + \chi^2}} \\
&= (1-k)\frac{N\beta^2}{\T\eta^2}\SQEQ.
\end{split}
\ee

Then,
\begin{multline}
\dot{\eta} = 3\left\{k\H_2\L_2\frac{\chi}{\T} - \bigl(N\H_1(k-1)^2+\H_2 k^2\bigr)\frac{\beta^2}{\T \eta^2}\right\}\times\\
\times\SQEQ,
\end{multline}
\noindent and
\be
\eta^2_{\rm eq} = \frac{\beta^2}{\N_2\chi k}\bigl\{N\H_1(k-1)^2 + \H_2 k^2\bigr\}.
\ee

Defining, for simplicity, 
\be
\mu \equiv \left(\frac{k-1}{k}\right)^{2/3}\frac{mN}{m_1}, \label{eq:mu}
\ee

\noindent we then find that $\dot{a}_1$ takes the simple form

\be
\boxed{\frac{\dot{a}_1}{a_1} = -\frac{\mu}{1+\mu} \frac{2\chi}{\T} \sqrt{\frac{5}{8}\frac{\chi}{k}\frac{1}{1+\mu} + \chi^2}. \label{exadot}}
\ee
\noindent This is the primary result of our work; in the next subsection we evaluate it given realistic assumptions for the protoplanetary disk conditions, and explore its implications.

\subsection{Planet migration rate due to ARM}

Recall that $\chi$ is a small parameter of order $0.001 - 0.01$ that quantifies the degree to which the gas disk deviates from a purely Keplerian flow. We can thus safely say that $\chi^2 \ll \frac{5}{8}\frac{\chi}{k}\frac{1}{1+\mu}$. (For our nominal value of $\chi = 0.005$, this hold so long as $\mu \ll 100/k$.) In that case, we may simplify Eq. (\ref{exadot}), and define the ARM rate of the planet as

\be
\frac{1}{t_\mathrm{ARM}}\equiv-\frac{\dot{a}_1}{a_1} \simeq \frac{\mu}{(1+\mu)^{3/2}} \frac{2\chi}{\T} \sqrt{\frac{5}{8}\frac{\chi}{k}} \sim \chi^{3/2}\frac{M_{\mathrm{swarm}}}{m_1}, \label{eq:arm}
\ee
\noindent where, to recap, $a_1$ is the semi-major axis of the planet, $\mu$ is proportional to the ratio of the particle swarm mass $M_{\mathrm{swarm}}$ (the total mass of debris trapped in resonance) to planet mass $m_1$ (Eq. \ref{eq:mu}), k is the resonance number, and $\T$ is the AD eccentricity damping rate (Eq. \ref{tau}) evaluated at the nominal resonance location, i.e. at the semi-major axis $[a]$ where $n([a])/n_1 = (k-1)/k$, $n$ being the mean motion. 

The dependence of the migration rate on $\mu$ can be understood intuitively as follows: the more particles there are, the more they act as a drain on the planet's orbital energy and increase the migration rate; thus $1/t_\mathrm{ARM} \propto \mu$. However, when the total planetesimal mass becomes comparable to the planet mass ($\mu \sim 1$), the faster the planet runs away from the planetesimals, the less deep into resonance they fall, achieving a smaller equilibrium eccentricity and reducing the effectiveness of AD in extracting energy from the planet; thus, it is not surprising, and indeed necessary, that $1/t_\mathrm{ARM}$ be inversely correlated with $(1+\mu)$. In consequence, the ARM rate is a non-monotonic function of $\mu$, and it can be shown that it peaks exactly at $\mu = 2$. Thus, the highest possible migration rate is achieved at
\be
\mu = 2,~{\rm or} \quad mN = 2 \left(\frac{k}{k-1}\right)^{2/3} m_1,
\ee

Evaluating Eq. \ref{eq:arm}, we get
\begin{align}
t_\mathrm{ARM} &= 167 ~\mathrm{kyr}~ \left(\frac{\mu}{2}\right)^{-1}\left(\frac{1+\mu}{3}\right)^{3/2}\left(\frac{a_1}{1~\mathrm{AU}}\right)^{5/2} \times \nonumber \\
&\times \left(\frac{\chi}{0.005}\right)^{-3/2}\left(\frac{\Sigma_0}{2000~\mathrm{g/cm^2}}\right)^{-1}\left(\frac{h/r}{0.05}\right) \times \label{tarm}\\
&\times \left(\frac{s}{1~\mathrm{km}}\right)\left(\frac{\rho}{2~\mathrm{g/cc}}\right)\left(\frac{M_\star}{M_\odot}\right)^{-1/2}\left(\frac{k}{3}\right)^{13/6}\left(\frac{2}{k-1}\right)^{5/3}, \nonumber
\end{align}
\noindent where $\Sigma_0$ is the gas surface density at $1$ AU, $h/r$ is the aspect ratio of the disk, and $\rho$ is the material density of a planetesimal. Note that the ARM timescale is linear in planetesimal size, and is a steep function of $a_1$. 

Figure \ref{analyticals} presents the time $t_{\mathrm{ARM},a0} \equiv \frac{2}{5}t_\mathrm{ARM}$ that it takes for a planet of mass $M_p$ starting at a semi-major axis $a_0$ to be pushed into the sun by a swarm of $1-$km planetesimals of mass $M_\mathrm{swarm}$ trapped in the $3:2$ resonance. As we can plainly see, this is an extremely efficient migration mechanism. While it can be argued that it is difficult to assemble several Earth masses' worth of $1-$km planetesimals, at $0.5$ AU even $10-$km particles can accomplish the job well within $1$ Myr (recall that the ARM timescale is linear in planetesimal size). 

We may now compare the ARM timescale to the standard type I migration scenario, though, as discussed previously, the magnitude and even direction of type I migration are highly uncertain. Nevertheless, we use the semi-analytical formulae of Izidoro et al. (2015) to compute the type I migration timescale. Since the ARM rate scales quite gently with the amount of mass in resonance, for definitiveness we use the minimum ARM timescale (Eq. \ref{tarm}), assuming $\mu = 2$ (this will be further justified by the results of Section III). Figure \ref{typeIcomp} presents a color plot of the ratio of the two timescales, as a function of planet mass and semi-major axis, for two different planetesimal sizes. For smaller planet masses, ARM easily dominates inside $1$ AU. For more massive planets, at $s=1$ km this boundary is pushed inwards to $\sim 0.5$ AU. 

To summarize, we have presented an analytical theory of Aero-Resonant Migration (ARM) and derived the migration timescale $t_\mathrm{ARM}$. Unsurprisingly, migration is most efficient when the mass of planetesimals in resonance and the mass of the planet are comparable; in such cases, the migration is extremely efficient: $s=1$ km planetesimals are able to push a planet from $1$ AU into its sun in just under $70$ kyr, with the timescale increasing linearly with planetesimal radius. Due to its steep dependence on semi-major axis ($t_\mathrm{ARM} \propto a^{5/2}_p$), ARM dominates over Type I migration in the inner regions of the disk, especially for smaller mass planets. 

We note that, up until this point, our analysis of ARM has been entirely analytical, and, as yet, untested. In the next section, we present a series of numerical N-body experiments that confirm that the analytical timescale presented in Eq. (\ref{tarm}) is both qualitatively and quantitatively accurate. 

\section{ARM: numerical experiments}

We use the N-body code $mercury6$ (Chambers 1999) to simulate the resonant interaction between a planet and a swarm of exterior planetesimals. To the standard $mercury6$ equations, we add a user-defined acceleration that applies the effect of Aerodynamic Drag to planetesimals (Eq. \ref{adrag}), calculated using the exact same disk parameters as those of Eq. (\ref{tarm}). Out of necessity, we employ the super-particle approximation, using a few hundred relatively-higher-mass particles as a proxy for a swarm of tiny planetesimals. The acceleration $\mathbf{a}_{\mathrm{drag}}$ applied to each particle is that of a $s=1$ km, $\rho=2$ g/cc planetesimal (since the AD timescale is linear in the planetesimal radius and density, migration timescales due to planetesimals of other sizes can be simply extrapolated). In addition, we damp the planet's eccentricity and inclination on timescales $t_e = t_\mathrm{wave}/0.78$ and $t_i = t_\mathrm{wave}/0.544$, respectively, where  
\be
t_\mathrm{wave} = \left(\frac{M_\odot}{M_p}\right)\left(\frac{M_\odot}{\Sigma_\mathrm{gas}a^2}\right)\left(\frac{h}{r}\right)^4 n^{-1},
\ee
\noindent as is standard for a planet embedded in a gaseous disk (e.g. Izidoro et al. 2015). 

\subsection{Migration timescale}

\begin{figure}
\centering
\scalebox{0.42}{\includegraphics{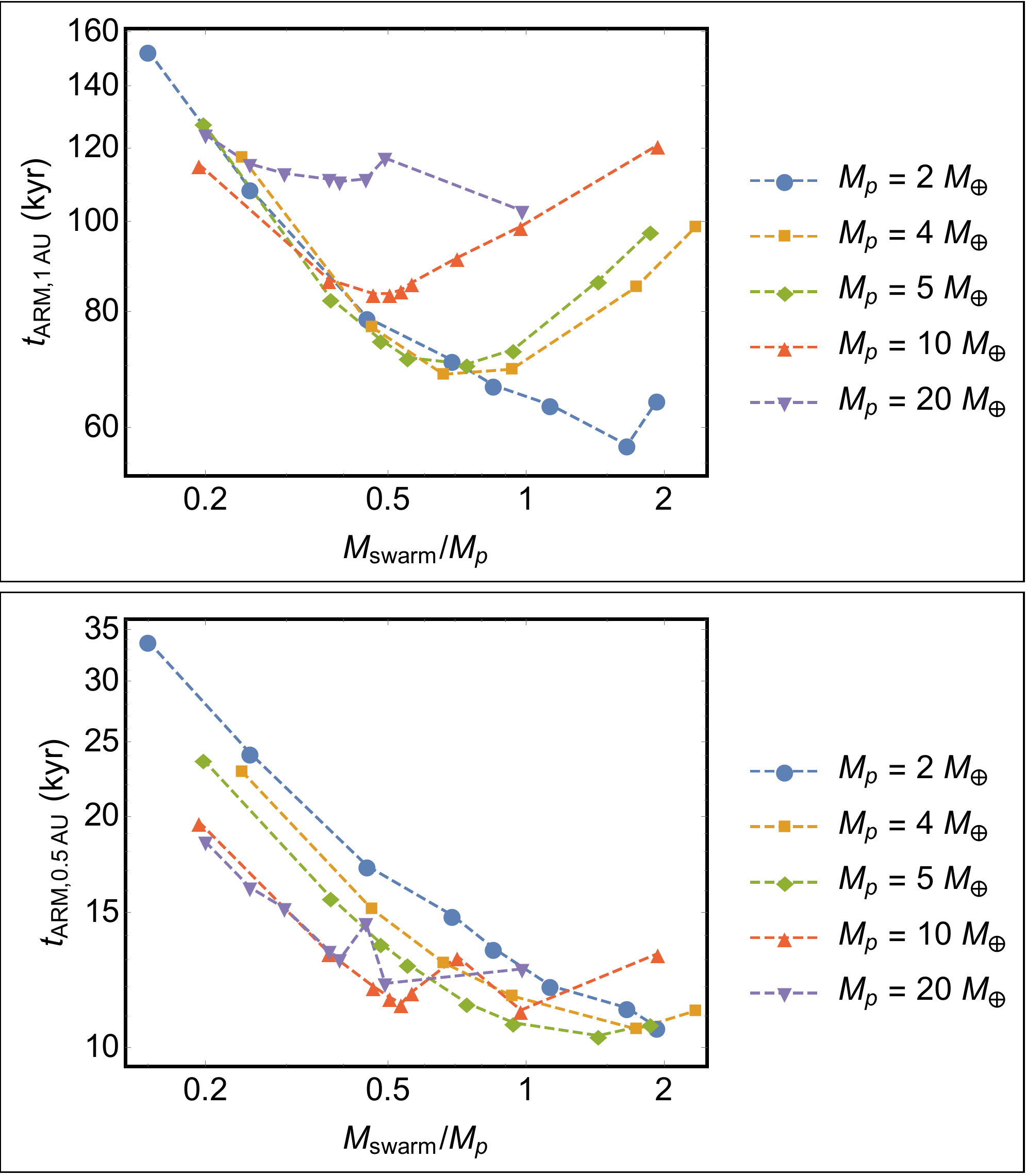}}
\caption{ARM timescales vs the ratio of total planetesimal mass $M_\mathrm{swarm}$ to planet mass $M_p$, based on numerical N-body integrations (see text for description). \textit{Top:} starting from $1$ AU. \textit{Bottom:} starting from $0.5$ AU.}
\label{numtimes}
\end{figure}

We initialize a planet (of varying mass) at $1$ AU, and a swarm of test particles (of varying total mass) just outside the nominal $3:2$ resonance location. The particles interact with the planet but not with each other. We integrate the system using the hybrid symplectic/Bulirsch-Stoer integrator with an initial timestep of $3$ days. We stop the integration when the planet reaches the inner edge of the disk, which we define to be at $a=0.1$ AU, and record the total migration time from $1$ AU, as well as the migration time from $0.5$ AU. Since our goal is primarily to get confirmation of the main analytical results of Section II, we do not carry out a full statistical treatment of the experiment; each combination of $M_p$ and $M_\mathrm{swarm}/M_p$ is simulated only once. Figure \ref{numtimes} presents the results. 

Figure \ref{numtimes} should be compared with Figure \ref{analyticals} of Section II. First, we note the qualitative similarity: just as predicted by our analytical calculation, the migration timescale has a minimum. The location of the minimum is not exactly as theorized (for the $3:2$ resonance, $\mu = 2$ would correspond to $M_\mathrm{swarm}=2.6 M_p$), and depends slightly on planet mass. In general, the location of the minimum appears to be more forgiving: less mass is required to attain the maximum migration rate than we supposed. Quantitatively, the numerical experiments are also in good agreement with the model: the minimum migration timescales are essentially identical, and a ten-fold variation in $M_\mathrm{swarm}/M_p$ results in only a mild, factor of $2-3$ change in the migration timescale. 

We thus conclude that the numerical experiments corroborate the analytical findings of Section II, exhibiting both qualitative and quantitative agreement. Having confirmed the mild dependence of the migration timescale on the particle to planet mass ratio, we also feel justified in using the minimum ARM timescale in our comparison of the ARM and type I timescales (Figure \ref{typeIcomp}). 

\subsection{Azimuthal distribution of planetesimals in resonance}

\begin{figure*}
\centering
\scalebox{0.45}{\includegraphics{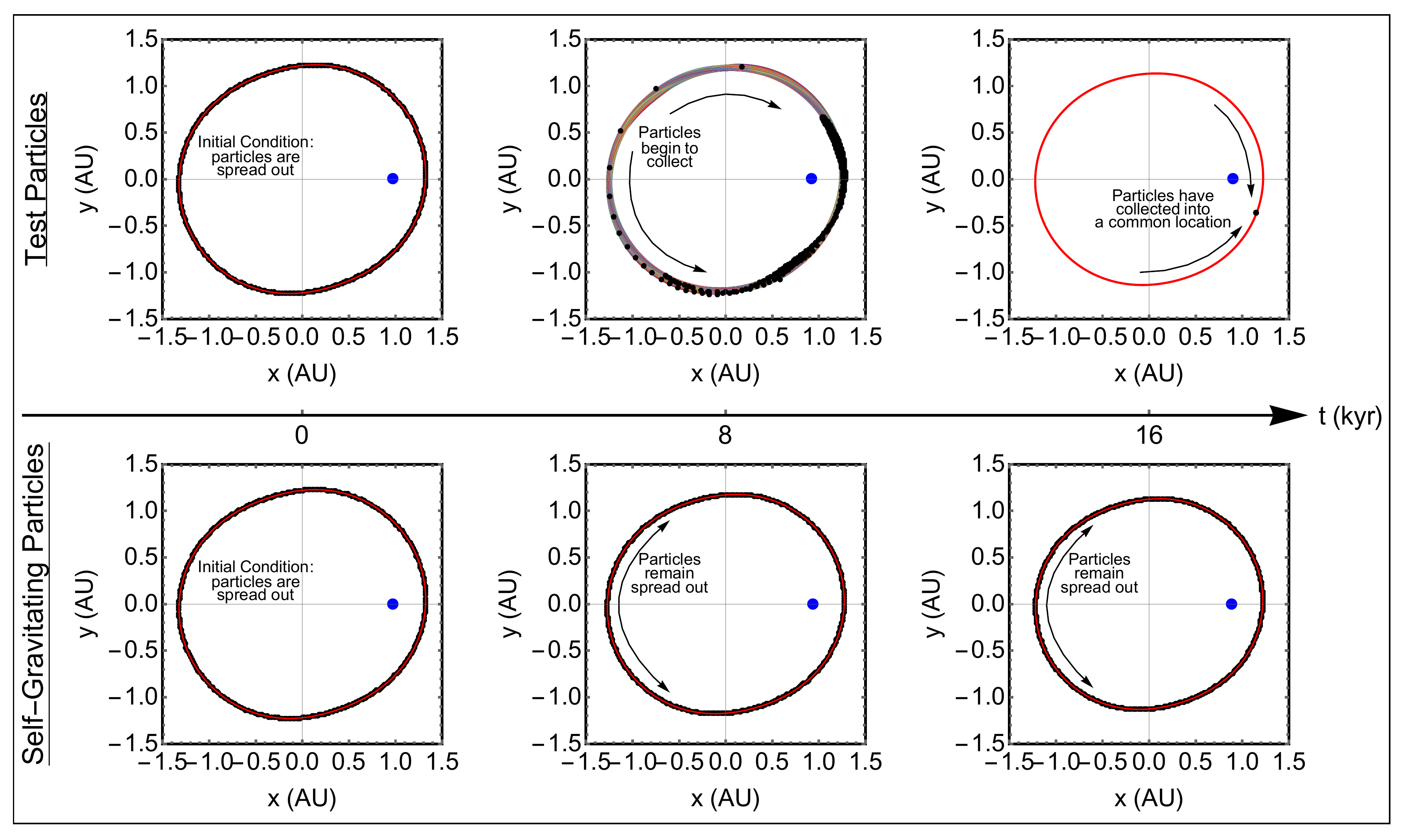}}
\caption{Three snapshots from two N-body integrations, one that treats planetesimals as test particles (top), and one that considers the self-gravitating interactions between them (bottom). In each case, $320$ particles (black points) totaling $2 M_\oplus$, are initialized in $3:2$ resonance with a $10 M_p$ planet (blue). Each snapshot is presented in a reference frame co-rotating with the planet. \textit{Left:} The particles are initially spread out along a single resonant orbit (red). The initial conditions for the two integrations are identical. \textit{Middle:} By $t=8$ kyr, the test particle orbits (top, shown in color) have become disordered and the test particles have begun to collect together. The self-gravitating particles remain spread out. \textit{Right:} By $t=16$ kyr, the test particles have collapsed to a single point (black) on a single resonant orbit (red), while the self-gravitating particles are still azimuthally spread out.}
\label{rotframe}
\end{figure*}

A trivial objection that can be envisioned against Aero-Resonant Migration is the following: if there are several Earth masses of planetesimals trapped in a resonance together, why do they not form a planet? After all, the ARM mechanism depends on the planetesimals remaining small and separate, so that they experience significant aerodynamic drag and essentially act as a ``parachute'' for the planet. Collapse into a planet of their own is, then, the most obvious failure mode.

Planet formation, by definition, requires self-interaction between particles. We therefore consider another numerical experiment, in which the planetesimals' mutual gravitational interactions are taken into account. Because we are employing the super-particle approximation, the result of this experiment is necessarily suspect, but we do believe that it can be, at the very least, suggestive. 

\begin{figure}
\centering
\scalebox{0.45}{\includegraphics{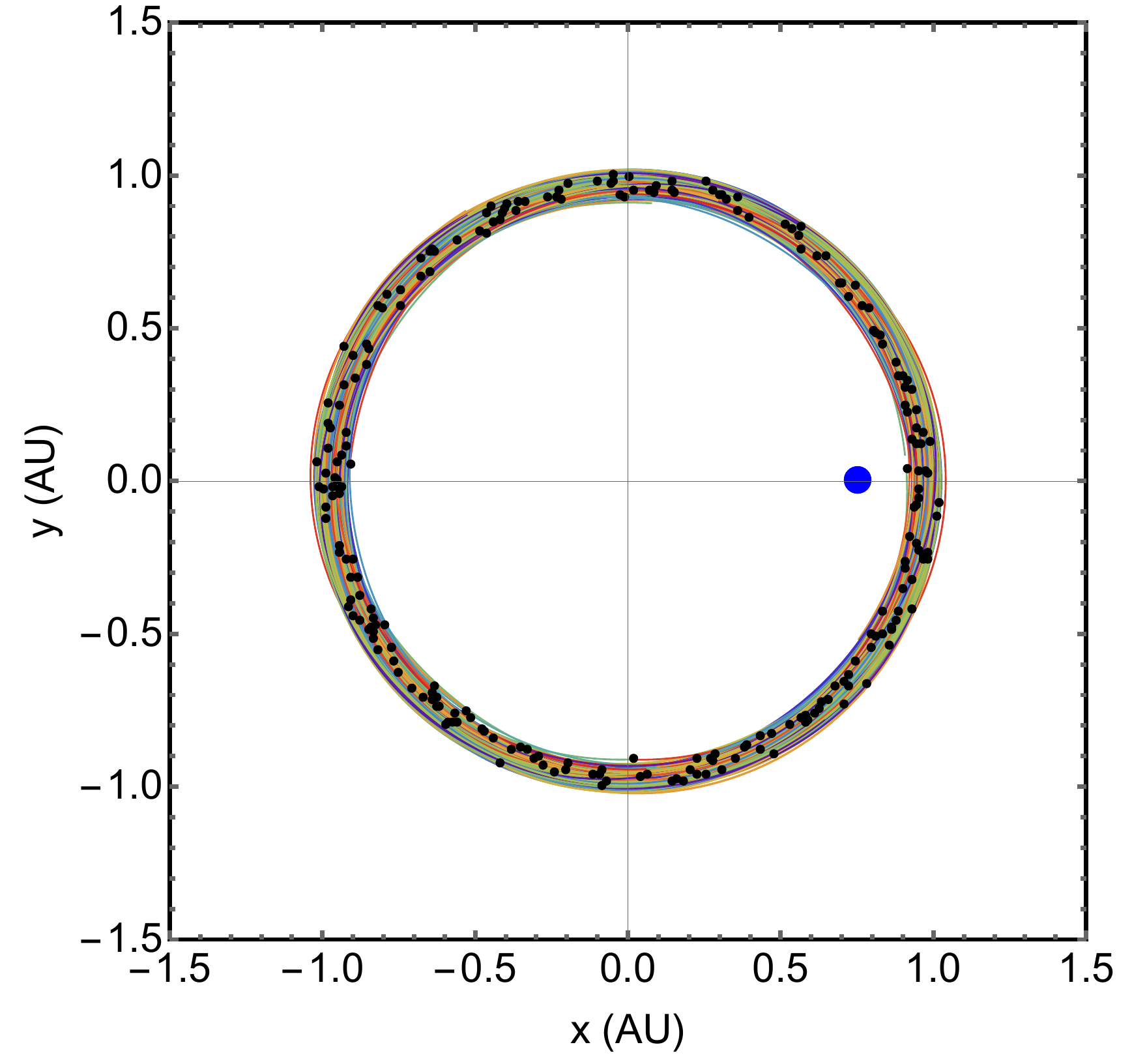}}
\caption{A later snapshot ($t=42$ kyr) of the N-body integration that takes into account mutual interaction between planetesimals (Fig. \ref{rotframe}, bottom), shown in the corotating frame of the planet (blue). Here, the particle orbits (various colors) have become more disordered, but the particles (black points) are still azimuthally spread.}
\label{rotframe2}
\end{figure}

We consider $320$ particles, with a total mass of $2 M_\oplus$, in $3:2$ resonance with a $10 M_\oplus$ planet. The particles are initially spread out azimuthally; in the frame of reference co-rotating with the planet's mean motion, they occupy a single resonant orbit but are each at a different phase. We then run two integrations: one which accounts for their self-interactions, and one that does not. Figure \ref{rotframe} presents three snapshots from the early stages of that integration. Perhaps somewhat counter-intuitively, self-gravitation of the particle swarm impedes conglomeration. That is, test particles collapse to a point relatively quickly; this is, essentially, because they raise the eccentricity of the planet, and then assume a state where both of their resonant angles are librating, which uniquely constrains both the shape and the phase of their orbits. On the other hand, in the integration that takes mutual particle interactions into account, no significant change in the particles' orbits have occurred. Evidently, and rather unexpectedly, in this case the mutual interactions have a stabilizing effect on the system. 

Throughout the subsequent integration, the mutually interacting particles do not exhibit any clumping behavior. Figure \ref{rotframe2} shows a later snapshot of the same N-body integration as that presented in Figure \ref{rotframe} (bottom). Here the particle orbits have relaxed into a more disordered state, occupying a thicker torus of space, yet the particles remain azimuthally dispersed. This state persists throughout the remainder of the integration. Due to this relaxation, we find that the migration timescale increases slightly, by a factor of about $\sim 1.4$. Assuming that this is typical, the conclusions of Section II are off by at most a factor of $\mathcal{O}(1)$, and thus remain valid. 

Note that, even though the self-interaction of particles appears to keep them azimuthally distributed, some particle collisions still inevitably happen. We treat these collisions as coagulations (the final product is a single particle with the same density as the colliding particles, and mass equal to the their total mass) in order to approximately preserve the total mass in resonance, even though they are not necessarily so. Of the initial $320$ particles, about $50$ remain by the time migration is halted at the disk's inner edge. We are not claiming, therefore, that collisional coagulation is entirely prevented; rather, we simply point out that self-interaction may prevent particles in resonance from collecting azimuthally and having their collision rates further enhanced in that way.

A complete and realistic treatment of the collisional interactions of particles inside the resonance is beyond the scope of this work. However, we note that bodies of radius $10^4-10^5$ cm are at the minimum of the catastrophic destruction curve (Leinhardt \& Stewart 2009) and only require a specific energy of $Q\sim 10^6-10^7$ ergs/g to continue their downward decay within the collisional cascade. Notably, if we adopt a velocity dispersion of $\sim 0.01 v_{\rm K}$ at $1$~AU, typical impacts between $\sim 1$ km bodies will yield $Q\sim (0.01 v_{\rm K})^2/2 \sim 5 \times 10^8$ ergs/g, well in excess of the catastrophic destruction curve. That is to say, it is unclear (and perhaps even unlikely) that collisional grinding within the resonance would lead to coagulation at all. A more relevant picture may be one where collisions allow planetesimals to exit the resonance and either get re-captured (if the change in semi-major axis is positive) or skip inwards to the next (higher-index) resonance (if the change in semi-major axis is negative). Naturally, this recycling of material cannot last forever, and there is no doubt that long-range aero-resonant migration necessitates a prolonged flux of external, inward-drifting debris. However, the key point here is that coagulation within resonance is far from an assured outcome of the ARM process.

In this work, we have not carried out a careful investigation of the analytic machinery that underlies the azimuthal randomization of resonant debris through self-interactions. Moreover, because we are severely limited in the number of self-interacting particles we can simulate, the particle masses we consider here are approximately half that of the moon, yet we apply aerodynamic drag to them as though they were $1$-km particles. We therefore cannot claim to have attained convergence in terms of resolution, and cannot rule out the possibility that the stabilizing effect we observe may be artificial in nature, caused by, e.g., the enhanced self-stirring the super-particles experience.

We speculate, however, that the effect is qualitatively more subtle. In particular, it is likely that due to the stochastic potential of the particles, the libration amplitudes of the resonant harmonics never reach zero, which prevents the planetary eccentricity from being coherently excited. Indeed, it is the fact that $e_p = 0$ that prevents the particles from coagulating, rather than chaotic scattering among debris. Thus, gravitational self-stirring need not actually randomize the particles directly -- all that is necessary is to stochastically enhance the phase-space area around the corotation resonance equilibrium point such that it engulfs the origin of the phase-space portrait. Moreover, because the resonant equilibrium eccentricity of the planet is close to zero anyway, only a mild perturbation is needed to set the resonant angle in circulation. Consequently, as long as $e_p = 0$, the results should hold. The conditions under which this state can be maintained are not obvious, however, and more detailed theoretical work is required in order to determine whether this tendency of self-interacting particles trapped in resonance \textit{not} to clump together is indeed physical.

We note that if this turns out not to be the case, a number of physical effects may be invoked instead. For example, turbulence in the disk may prevent the planetesimal congregation through a similar process. Alternatively, external collisions may cause a proto-protoplanet to break up again. Or the particles may encounter each other at speeds high enough to be destructive rather than constructive. Yet another possibility is that as larger particles are built up, they may escape the resonance, and be replaced by fresh AD-driven inflowing debris. At worst, the efficacy and rate of Aero-Resonant Migration may become uncertain, possibly subject to a few tunable parameters. 

\section{Conclusion}

In this work, we have developed the theory of ``Aero-Resonant Migration'' (ARM), a smooth disk migration mechanism for small planets ($2 M_\oplus \lo M_p \lo 30 M_\oplus$) in which small planetesimals ($10 \mathrm{m} \lo s \lo 10$km) undergo orbital decay due to aerodynamic drag and resonantly shepherd planets ahead of them. In approaching this problem, we have purposely avoided the specifics of any particular scenario that might lead to ARM. Instead, our aim has been simply to characterize the relevant physical process from analytical and numerical grounds. 

We find ARM to be a robust migration mechanism, able to operate in a large section of parameter space. While the maximum migration rate requires that the mass of planetesimals in resonance be similar to that of the planet, the scaling is forgiving; $2 M_\oplus$ of planetesimals can shepherd a $10 M_\oplus$ planet nearly as effectively as $10 M_\oplus$ of planetesimals. Due to its steep dependence on semi-major axis, ARM operates most efficiently in the inner part of the protoplanetary disk ($\lo 0.5$ AU). In this region, for small planet masses, ARM's efficiency is similar to that of type I migration. 

We find that self-interaction between the planetesimals trapped in resonance may play a surprisingly important role in facilitating aero-resonant migration. In our numerical experiments -- which admittedly employ the super-particle approximation and are thus merely suggestive -- we have discovered that, contrary to what may be naively expected, self-interacting particles trapped in resonance do not tend to clump together. Instead, they remain azimuthally spread out, thus defeating the most obvious objection to ARM -- to wit, if there are several Earth masses of solids trapped in a resonance, why do they not form a planet? Self-interaction, it seems, may be the key to the answer. 

Moving forward, several avenues of research relating to ARM, and applications thereof, may be pursued. The most obvious of these is further inquiry into the above-mentioned effect of self-interaction. Hot on its heels is the question of how much particle mass can be realistically accumulated in a resonance, on what timescales, and what distribution of sizes the particles would have -- and how these considerations affect the efficiency of ARM. On the application front, perhaps the most exciting puzzle is that of ultra-short-period planets. These planets reside in the magnetospheric cavity of their host star, and thus likely could not have formed there. However, they could have been driven there by a process related to ARM during episodic peaks of accretion rates of the host stars. 

To conclude, given an appreciable mass of small planetesimals in resonance, the ARM timescale is short enough that this mechanism is capable of driving significant planet migration during the lifetime of the protoplanetary disk. While more work is required to assess the particulars of the process, such as the importance of self-interaction between particles trapped in resonance, it is nevertheless clear that ARM may play an important role in shaping planetary system architectures.

\section*{Acknowledgments}
We are thankful to Alessandro Morbidelli, Sean Raymond and Greg Laughlin for illuminating discussions, as well as to the anonymous referee for providing a thorough and insightful report. NIS gratefully acknowledges support through the Sherman Fairchild Fellowship at Caltech. KB gratefully acknowledges the David and Lucile Packard Foundation and the Alfred P. Sloan Foundation for their generous support.

%%%%%%%%%%%%%%%%%%%%%%%%%%%%%%%%%
\end{document}